\documentstyle[12pt,epsf,subfigure]{article}
\def \blankline{\vspace{0.4 cm}}
\newcommand{\PSbox}[3]{\mbox{\rule{0in}{#3}\includegraphics{#1}\hspace{#2}}}
\begin {document}
\centerline {\large \bf A 10 MHz Beam Counter and a Multiplicity Detector }
\centerline {\large \bf for the E864 Spectrometer }
\blankline
\vskip 1.0cm
\centerline { P. Haridas$^{a}$, I.A. Pless$^{a}$,  G. Van Buren$^{a}$, 
J. Tomasi$^{a}$,M.S.Z. Rabin$^{b}$, }
\centerline { K. Barish$^{c}$, R.D. Majka$^{c}$ }
\vskip 1.0cm
\leftline{a. Massachusetts Institute of Technology, Cambridge, MA 02139, USA.}

\leftline{b. University of Massachusetts, Amherst, MA 01003 USA.}

\leftline{c. Yale University, New Haven, CT 06520 USA.}

\vskip 1.0cm
\centerline{\bf{Abstract}}
\vskip 0.5cm

The E864 experiment at BNL requires a beam counter and 
multiplicity detector system
that can perform at an incident beam rate of 10$^{7}$ Au ions per second. 
We have developed and tested a
150~$\mu$m thick quartz Cherenkov beam counter and a scintillator
based multiplicity-trigger counter during the first run of this experiment
in 1994. We obtained a time resolution of 78~ps for the beam counter at an
incident beam rate 5$\times$10$^{5}$~Hz
and 100~ps at a rate of 1$\times$10$^{7}$~Hz.
Pulse height discrimination is used to obtain a minimum bias and a 10\%
centrality trigger from the multiplicity detectors. The multiplicity
counter has a time resolution of 250~ps.

\vskip 1.0cm

\noindent{\large \bf 1. Introduction} 

\blankline

   We report on a beam and 
multiplicity-trigger counter system
that we have developed and tested for the E864 spectrometer 
at the Brookhaven National Laboratory (BNL).
 E864 is designed to search for rare composite states, stranglets and 
other novel forms of matter which may be produced in
a small fraction of relativistic Au-Pb collisions at the BNL 
Alternating Gradient
Synchrotron (AGS). The momentum of the Au beam at the AGS is 11.7~A~GeV/c.
The beam, veto and multiplicity counters constitute the
front end of the E864 spectrometer.
This spectrometer provides momentum, time of flight and
energy (calorimetry) information for charged and neutral secondary particles 
produced in central ( small impact parameter ) Au-Pb interactions.  
In order to realize a sensitivity of 
10$^{-11}$ per interaction for the detection of rare particles 
within a reasonable running period, 
it is required that the incident
beam intensity be of the order of 10$^{7}$ s$^{-1}$.  Stable long term 
performance and good time resolution (80 - 100 ps) at these 
high rates were the central
criteria that dictated the design and development of the beam and trigger
counters.

\blankline

	As is well known, light emission of a solid scintillator rapidly
deteriorates due to radiation damage when exposed to heavy-ion beams. To
minimize radiation damage as well as 
the number of beam particle interactions in the detector
material we have used a thin (150~$\mu$m) quartz plate for the beam counter.  
Cherenkov light emitted by a relativistic ion traversing the quartz plate
emerges through the edges of the plate after a series of total internal 
reflections.  Two photomultipliers (PMT)
optically coupled to opposite edges of the 
quartz plate provide the beam counter with superior signal characteristics as
compared to a scintillator counter of similar construction.  Although this 
technique has been employed in other experiments [1],
the reported results are at an incident beam intensity of about 10$^{6}$~Hz,
which is an order of magnitude less than the desired E864 beam intensity 
of 10$^{7}$~Hz. Thus the main
effort behind the development of our beam counters has been to achieve good
time resolution at high rates while maintaining small PMT anode currents to 
prevent the deterioration of the photomultiplier tubes due to overheating of
the dynodes.
In Section 2 we describe the
construction and signal characteristics of the beam counter and in Section 4 
we give results on time resolution and stability.

\blankline

     The E864 experiment uses a multiplicity trigger to select central
events, because in relativistic heavy-ion collisions
there is a strong correlation 
between particle multiplicity and centrality.  The E864 multiplicity detector
is an annular piece of fast BC420  scintillator placed around the beam pipe
13~cm downstream of the target.  The annulus is segmented into four 
quadrants and each quadrant is viewed by a photomultiplier.  The centrality
trigger is obtained by imposing a high discriminator threshold on the
summed signals from the four counters.  We discuss the multiplicity counter in 
detail, in Section 3.

\blankline

	Fig.~1 shows the entire beam and multiplicity-trigger counter 
assembly. The quartz Cherenkov beam counter is housed in a light tight box
that is located on the extreme left side in Fig.1. In order to veto beam
particles which are more than 0.5 cm away from the beam axis 
a quartz-Cherenkov hole 
counter is placed immediately upstream of the beam counter, within the same 
light-tight enclosure housing the beam counter. Another, larger hole veto
counter made of scintillator is placed downstream of the beam counter to
detect and veto those events in which a beam particle interacted in the 
beam counter quartz plate. This counter also helps to veto events 
which have a beam particle 
accompanied by a halo of charged particles that originate from interactions
in the upstream beam magnets and collimators. The target assembly and the 
multiplicity-trigger counters are shown on the right side in Fig.1.
Lead shielding placed between the target and the scintillator hole counter
minimizes the possibility of triggering the veto counter by secondary particles
backscattered from the target. Lead shielding upstream of the scintillator
veto counter protects it from knock-on electrons ( $\delta$-rays ) produced 
by the gold beam passing through the upstream apparatus.

\blankline
\blankline

\noindent{\large \bf 2. The Quartz Beam and Hole Counter}
\vskip 0.3cm
\noindent{\bf 2.1 Detector Assembly}
\vskip 0.2cm
As mentioned previously, the quartz Cherenkov hole-veto and beam counters
are housed in a light-tight aluminium box with easily removable front  
and back cover plates for quick access. The photomultiplier tubes are 
mounted vertically with respect to the beam axis.
To describe the optical
part of the assembly in greater detail we provide in Fig.~2 a sketch of the two
detectors from side~(a) and perspective~(b) views, where VC and BC are the 
hole-veto and beam counters, respectively. The quartz 
plates for both counters are mounted in PET (Polyester Terephthalate) frames.
This material was chosen so that humidity changes will not affect the 
dimensional stability of the frame.
Two 0.5~cm thick lucite plates 
embedded in the PET frame provide the optical coupling between the quartz plates
and the front faces of the 2~inch phototubes.
The top and bottom edges of the beam counter quartz plate and one edge
of each of the veto counter quartz plates
are held in narrow slits on the front face (i.e. away from the PMT face)
of the lucite coupler.

\blankline

  Because the beam counter quartz plate (5~cm x 8.8~cm) is 
only 150~$\mu$m thick it is mounted in a way that minimizes the stress
on the plate and simultaneously ensures excellent optical contact with
the front face of the PMT. 
This is done by filling the slits containing the edges of the 
quartz plate with immersion oil (refractive index =1.515 ) as
illustrated in the 
expanded part in Fig.2.
The immersion oil is held in 
the slit by the capillary action of the liquid between the lucite and quartz
plate surfaces.  No dripping or draining of the oil was observed, regardless 
of the orientation of the frame. We chose Supersil-1 for the beam counter
quartz plate because of its low radiation damage properties [2], and its
property of being striation free in all three directions. While mounting
the plate, sufficient care has to be taken to ensure that no air bubbles are 
trapped in the immersion oil filling the slits.  The width of the slit proved to
be an important factor in achieving a bubble free quartz - lucite interface,
and a width of 375~$\mu$m was used in the final design.

	The hole veto counter uses a 1 mm thick Corning 2940 quartz plate.  
Because they are less fragile than the beam counter plate, these plates 
were optically cemented into the slits in the lucite.  
The hole veto counter consists of
two "L-shaped" pieces, which when mounted with a small overlap as
shown in Fig.~2(b), provide a center hole of dimension 1~cm x 1~cm. As each
quartz plate of the quartz hole counter is viewed by only one phototube,
the quartz plates need not be perpendicular to the beam. In fact, tilting
the quartz plate by 10 degrees maximized the amount of Cherenkov light 
that reaches each phototube.

\blankline

\noindent{\bf 2.2  Photomultiplier Tube Base modifications}

\blankline

  We have used Hamamatsu 12-stage photomultiplier tubes R4001 and R2059 
for the hole and beam counters,
respectively.  These PMTs belong to the 1828-01
family of 2~inch phototubes with a rise time of 1.4 ns and transit time jitter of
0.6 ns [3] at the recommended operating voltage of 2500~V.
These photomultiplier tubes were the most economical choice with adequate
rise time and a sufficiently high anode current limit.  
It is important for  
high rate (10$^{7}$~Hz) operation of 
PMTs to prevent voltage sag in the last dynodes, because such voltage sag will
cause a degradation of the rise time of the signal. It is also
important to maintain low PMT anode currents to prevent gain losses due to the
overheating of the last dynodes.  To satisfy both of these requirements
we must
operate the tubes at a low main voltage ($\sim$ 1600 V), with high current
voltage supplies on the eleventh and twelveth dynodes.  
This requires a modification of
the base resistor chain as shown in Fig.~3.  In addition to the external voltage
supplies, we have also connected 4.7 nF capacitators 
to all the dynodes
to provide a filter for high freqency noise.

\blankline

\noindent{\bf 2.3  Signal and Pulse Height Distribution}

\blankline

      Figure~4 shows the pulses from the top and bottom PMTs of the 
beam counter observed on a 300 MHz analog 
scope during a $\sim$1s spill of 11.7~A~GeV/c Au ions at a beam rate of 
1.2 x 10$^{7}$~Hz. The pulses have a rise time of about 2 ns, an average pulse
height of about 30mV (50$\Omega$ termination) 
and FWHM of about 4 ns. These pulses
were obtained after the original signals from the PMTs were split equally into
two, and thus represent 50\% of the original signals. 
Comparisons of these
pulses with those at lower rates ( 5$\times$10$^{5}$~Hz ) show no noticeble 
changes in the pulse height or pulse width. 
At an incident beam rate of 5 x 10$^{5}$ Hz, 
we estimate the anode current to be about 4$\mu$A.
At beam rates of 1 x 10$^{7}$ Hz the anode current is about 80$\mu$A, well 
below the 200$\mu$A maximum anode current limit set by the manufacturers.  
We also note here, that for AGS beams of 1 x 10$^{7}$ Hz, the 
"average continuous anode" current for each beam counter photomultiplier
is about 20$\mu$A, because the beam spills are 
at 4 sec intervals. By maintaining a low pulse height we have been
able to operate these PMTs at sufficiently low anode currents, to ensure
a longer life time for these photomultiplier tubes. We have also compared the
pulses at the begining and end of each spill during a high rate run
and seen no pulse height variations. Any observed differences would
suggest that the dynode voltages at the end of the resistor chain are sagging
below their preset voltages, or the presence of a dynode heating effect.

\blankline

   The pulse height spectrum as measured using a 
Fastbus ADC (LeCroy 1881) is shown
in Fig.~5. The least count for this ADC is 0.1~pC. The main peak of the 
pulse height spectrum is around ADC channel
850  and represents roughly a 40mV pulse that has been amplified 10
times, before being routed to the ADC input. This agrees quite well with
the pulse height as seen in Fig.4, where the peak height of the outer 
envelope of the 
band is around 40 mV. The second smaller peak around 1700 can be
identified as those events with 2 beam particles hitting the counter
simultaneously. Third and fourth peaks, corresponding to the simultaneous
passage of three and four gold ions, respectively, through the quartz plate 
are present at 2550 and 3400 ADC channels. A Gaussian fit to the main peak
gives a sigma of 102 channels which represents a 12\% width for the pulse height
distribution. 

\blankline

\noindent{\bf 2.4 Radiation Damage of the Quartz Plate}

\blankline
 
 Although the beam counter was tested at high rates (1$\times$10$^{7}$~Hz)
during the 1994 run, most of the data taking was performed at a beam rate
of 5$\times$10$^{5}$~Hz, for a total exposure of 10$^{11}$ Au ions. Comparison
of pulses taken at the begining and the end of the running period showed
very little variation in the pulse height, thus suggesting little or no
radiation damage to the quartz plate. However, during the 1995 run a gradual
degradation of the pulse height was observed as the quartz plate was exposed
to a beam rate of 5$\times$10$^{6}$Hz for extended periods of data taking.
The cause of this drop in pulse height was identified to be the acumulation of
a thin film of material in and around the region where the beam passed through
the quartz plate. This surface contamination of the quartz plate adversely
affects the total internal reflection of the Cherenkov light thus resulting
in a loss of pulse height. It was found that the plate could be thoroughly
cleaned using methanol, thus removing this surface deposition. Spectroscopic
studies showed the surface film to be made of carbon and oxygen compounds.     

\blankline

It should be noted here that all components of the beam counter including the
quartz plate was exposed to air. During the rest of the 1995 run, Nitrogen
was circulated through the beam counter housing in an attempt to prevent
the surface contamination of the quartz plate. Although this could
have reduced the rate of accumulation, close examination of the quartz plate at
the end of the run showed a faint spot on the quartz plate. For the forthcoming
run in 1996 we are redesigning the quartz plate holder so that it can be
operated in vaccuum. Additional design modifications are also being implemented
to facilitate quick changes of the quartz plate, if required, during the run.
 
\blankline

\blankline

\noindent{\large \bf 3. The Multiplicity Detector}

\blankline

\noindent{\bf 3.1 Detector Assembly}

\blankline

   The E864 multiplicity detector covers the angular range from 16.6
$^\circ$ to 45$^\circ$ ($\eta$ = pseudo rapidity = 1.92 $\rightarrow$ 0.88) 
measured with respect to the beam axis. It is constructed
out of four 1 cm thick scintillation counters that fit around the beam pipe as
shown in Fig.~6. A sectional view along the beam is shown in Fig.~1.  It is
located 13 cm downstream of the target and is tipped at an angle of 8$^\circ$
to the vertical. Simulation studies have
shown that as the beam traverses the target,   
$\delta$-ray production would result in about 200~MeV being deposited 
in the scintillation counter. To remove this $\delta$-ray effect we 
have shielded the entire active surface of the scintillator by a $\sim$ 6 cm 
thick Heavymet [4] ring, around the beam pipe.  As shown in Fig.~1 the counters 
are mounted on two moving stages that can slide towards or away from the beam 
pipe.  Each counter is made of a fast (BC 420) quadrant shaped piece of 
scintillator viewed by a 2 inch Hamamatsu 1828-01 phototube with the same PMT 
base as described in Section 2.  Booster voltages to the last two dynodes can 
be provided for high rate operation.  Each PMT is equipped with an LED
and a fiber optic channel so that long term behavior of the phototubes could
be monitored in an independent manner. The operating voltage of the PMTs
was set around 1600~V giving a 300~mV pulse height per counter for 
high multiplicity events. The signal from each PMT was first split into
two equal half-signals. Four half-signals from the four counters were
combined using a Phillip Scientific 744 Fan-in/Fan-out module to generate
a summed signal and this signal was discriminated to generate the 
three triggers (see next section ) used for the fall 1994 run.  

\blankline

\noindent{\bf 3.2 Trigger Scheme}

\blankline
 
Three different interaction triggers were developed for the E864 experiment.
A mimimum bias trigger (INT0)
required that the sum of the pulses from the 
four multiplicity counters had a threshold greater than 10 mV. 
A centrality trigger (INT2) was implemented by selecting the events
which had summed signals from all four counters greater than 340mV. 
This trigger selected those events in the highest 10\% of the multiplicity
distribution.
In addition, an intermediate multiplicity
trigger (INT1) was obtained by triggering on the 
summed signals above a 200 mV threshold.
Count down scalers were used to select some fraction of the 
minimum bias (INT0) and intermediate multiplicity events (INT1). All central
event triggers (INT2) were recorded. A plot of the pulse height distribution
of the minimum bias trigger is given in Fig.~7. The histogram represented by
the dotted lines represents the 10\% central events.   

\blankline

\noindent{\bf 3.3 Comparison with Monte Carlo Studies }

\blankline

Because of the correlation between particle mulitiplicity and centrality
a cut in the ADC value of the multiplicity counter translates into a 
selection of the impact parameter. 
To study the correlation between ADC value and 
centrality a Monte Carlo program was written. An event is simulated with
GEANT~[5] by sending a 11.71 GeV per nucleon gold ion 
through a 10\% lead 
target. This ion produces $\delta$-rays until it interacts at a
random depth in the lead target. This collision is simulated with the 
HIJET~[6] event generator. 
The geometry for the study included the target,
vacuum chamber, Heavymet shielding, and the scintillator multiplicity counter.
All of GEANT's physics processes were turned on for the simulation.
The program simulates the number of photoelectrons expected 
after taking into account
the energy deposited in the scintillator, the emission spectrum for the 
scintillator, the geometry of the detector, and the photocathode spectral 
response of the phototube. The photoelecron distribution from the 
simulated data appeared similar to the observed pulse height spectra from
the experiment.

\blankline

The results of the Monte Carlo study on the centrality selection of 
the multiplicity
counter can be seen in Fig.~8. Fig.~8(a) shows the distribution 
of impact parameters for three different cuts on the energy 
deposited in the multiplicity counter
as obtained from the program. Fig.~8(b) shows the trigger 
probability for different impact parameters for different cuts on the energy 
deposited in the multiplicity counter as obtained from the program.  The solid 
line represents a cut on the events with the 10\% most energy deposited; the 
dashed line represents the 5\% most energy deposited; 
the dotted line represents 
the 1\% most energy deposited. 

Fig.~8 demonstrates that the multiplicity counter can be used
to roughly select the centrality of an event. For example, if we make a cut on
events with the 10\% highest pulse heights,
we are selecting events with impact parameters corresponding to the smallest
10\% of the impact parameter distribution that one would expect from a 
geometric impact parameter distribution 88\% of the time. 
Even higher pulse height
cuts result in higher centrality events, although with very high cuts the 
acceptance for these high centrality events also goes down.

\blankline
\blankline

\noindent{\large \bf 4. Time Resolution and Stability}
   
\blankline

    The times of arrival of pulses from the counters were digitized with
Fastbus TDC modules (

Lecroy 1872A) with a least count of 50 ps. The pulse heights for the
beam counter pulses were 30-40 mV [7] and those signals were discriminated at a
threshold level of 10 mV. Using the ADC information from each PMT we have
slew corrected the TDC data independently for each counter. The slew
corrected distribution of the time difference of the two counters 
(t$_{B}$-t$_{A}$), at an incident beam rate of 10$^{7}$~Hz is given in Fig.9.
The $\sigma$ for this distribution as obtained by a Gaussian fit 
is 200~ps. From this value we obtain a single counter resolution
of 141~ps, and an overall resolution of the two counter system of 100~ps,
which is the uncertainity in the start time, t$_{0}$ ( = 0.5(t$_{A}$+t$_{B}$)) 
for the E864
experiment. At the lower beam rate of 5$\times$10$^{5}$ Hz our slew corrected 
time resolution is 78 ps.

  The time of arrival of the pulses from the four multiplicity counters
were also digitized using Fastbus TDC channels and each counter was slew
corrected. We have defined the time as measured by the multiplicity counter
to be the four-fold average of the individual times from the four counters.
To obtain the time resolution of the multiplicity counter we give in Fig.~10
the distribution of the average time minus the start time as measured 
by the beam counter. When fitted to a Gaussian this distribution gives a 
$\sigma$ of 250~ps which is the time resolution for our multiplicity
counter.

\blankline

\noindent{\large \bf 5. Summary}

\blankline

   We have developed and tested a beam counter and multiplicity detector
system at
an incident beam rate of 10$^{7}$ Au ions s$^{-1}$. To our knowledge (based
on published results) this is the first time a quartz plate as thin as 
150$\mu$m has been successfully tested for a heavy ion beam counter. The 
beam counter has a time resolution of 100~ps at an incident beam rate
of 1$\times$10$^{7}$ Hz and an energy resolution of 12\%. An important 
feature of this beam counter is the low average photomultiplier
anode current (80 $\mu$A) drawn during a one second spill of 10$^{7}$ Au ions.
The scintillator multiplicity counter provides a minimum-bias, intermediate
multiplicity and very high multiplicity (the highest 10\% of the multiplicity 
distribution) triggers.
The multiplicity counter has an overall time resolution of 250~ps.
The beam counter and multiplicity detector system show 
good stablity as the incident
beam rate is varied from 10$^{5}$~Hz to 10$^{7}$~Hz.

\section*{Acknowledgements}

This work was supported by grants from the Department of Energy (DOE) High
Energy Division, the DOE Nuclear Division and the machine shop support
of the Physics department at the University of Massachusetts, Amherst.
   We would like to thank machinists Bob Verner and Richard Wilkey of the
University of Massachusetts, Amherst for an excellent job of building
the beam and multiplicity counters. We also thank engineers Will Emmet
and John Sinnott (Yale University) and Bill McGahern (BNL) for their design 
of the E864 front-end support systems and
for their assistance in the final installation of the counters in the 
BNL beamline.


\newpage
\centerline{\large \bf Figure Captions}

\blankline

\leftline{Figure 1. The E864 front-end configuration of detectors consisting of the}
\leftline{~~~~~~~~~~~~~Quartz-Cherenkov hole-veto and beam counters on the left, a}
\leftline{~~~~~~~~~~~~~scintillator hole-veto counter in the middle, followed by the}
\leftline{~~~~~~~~~~~~~target assembly and Multiplicity-trigger counters on the right.} 
\vskip 0.4cm
\leftline{Figure 2. A sketch illustrating the optical assembly of the quartz hole-veto}
\leftline{~~~~~~~~~~~~~and beam counters; (a) shows the side view and (b) the perspective}
\leftline{~~~~~~~~~~~~~view. }
\vskip 0.4cm
\leftline{Figure 3. The modified base for the Hamamatsu 1828-01 PMT with power }
\leftline{~~~~~~~~~~~~~suplies attached to the last two dynodes }
\vskip 0.4cm
\leftline{Figure 4. Beam counter pulses during a spill of
 1$\times$10$^{7}$ Au ions. The vertical }
\leftline{~~~~~~~~~~~~~~scale is 20~mV/div and the horizontal scale 
is 5~ns/div.}
\vskip 0.4cm
\leftline{Figure 5. The pulse height spectrum of the beam counter at an
incident }
\leftline{~~~~~~~~~~~~~~beam rate of 1.2$\times$10$^{7}$ Au ions per second.}
\vskip 0.4cm
\leftline{Figure 6. Beam's eye view of the multiplicity-trigger counters.}
\vskip 0.4cm
\leftline{Figure 7. The averaged (over the four counters) multiplicity detector pulse}
\leftline{~~~~~~~~~~~~~~height distribution for minimum-bias (solid line) and 10\%}
\leftline{~~~~~~~~~~~~~~most central events (dashed line).}
\vskip 0.4cm
\leftline{Figure 8. Monte Carlo study of the centrality selection of the multiplicity counter.}
\leftline{~~~~~~~~~~~~~~The three different curves represent different cuts on the energy }
\leftline{~~~~~~~~~~~~~~deposited in the multiplicity counter. The solid line represents a cut} 
\leftline{~~~~~~~~~~~~~~on the events with the 10\% most energy deposited, the dashed line } 
\leftline{~~~~~~~~~~~~~~represents the 5\% most energy deposited, and the dotted line represents } 
\leftline{~~~~~~~~~~~~~~the 1\% most energy deposited. (a) Distribution of impact parameters }
\leftline{~~~~~~~~~~~~~~(in fm) which pass a 10\%, 5\% and 1\% cut on the energy deposited in }
\leftline{~~~~~~~~~~~~~~the four scintillation counters in the multiplicity detector. }
\leftline{~~~~~~~~~~~~~~(b) Trigger probability for a particular impact parameter event to pass}
\leftline{~~~~~~~~~~~~~~a 10\%, 5\%, and 1\% energy deposit cut.}
\vskip 0.4cm
\leftline{Figure 9. The distribution of the time difference between the bottom (B) and top (A)}
\leftline{~~~~~~~~~~~~~~PMTs of the beam counter. $\sigma$(t$_{B}$-t$_{A}$) = 200~ps and }
\leftline{~~~~~~~~~~~~~~$\sigma$(t$_{0}$)=$\sigma$(0.5(t$_{B}$+t$_{A}$)) = 100~ps.}
\vskip 0.4cm
\leftline{Figure 10. The averaged (over the four counters) intrinsic time resolution ($\sigma$=250~ps) }
\leftline{~~~~~~~~~~~~~~of the multiplicity-trigger counter system.}

\newpage
\begin{figure}[htb]
\begin{center}
\vskip 1.5 cm
\mbox{\epsfysize=12.cm\epsffile{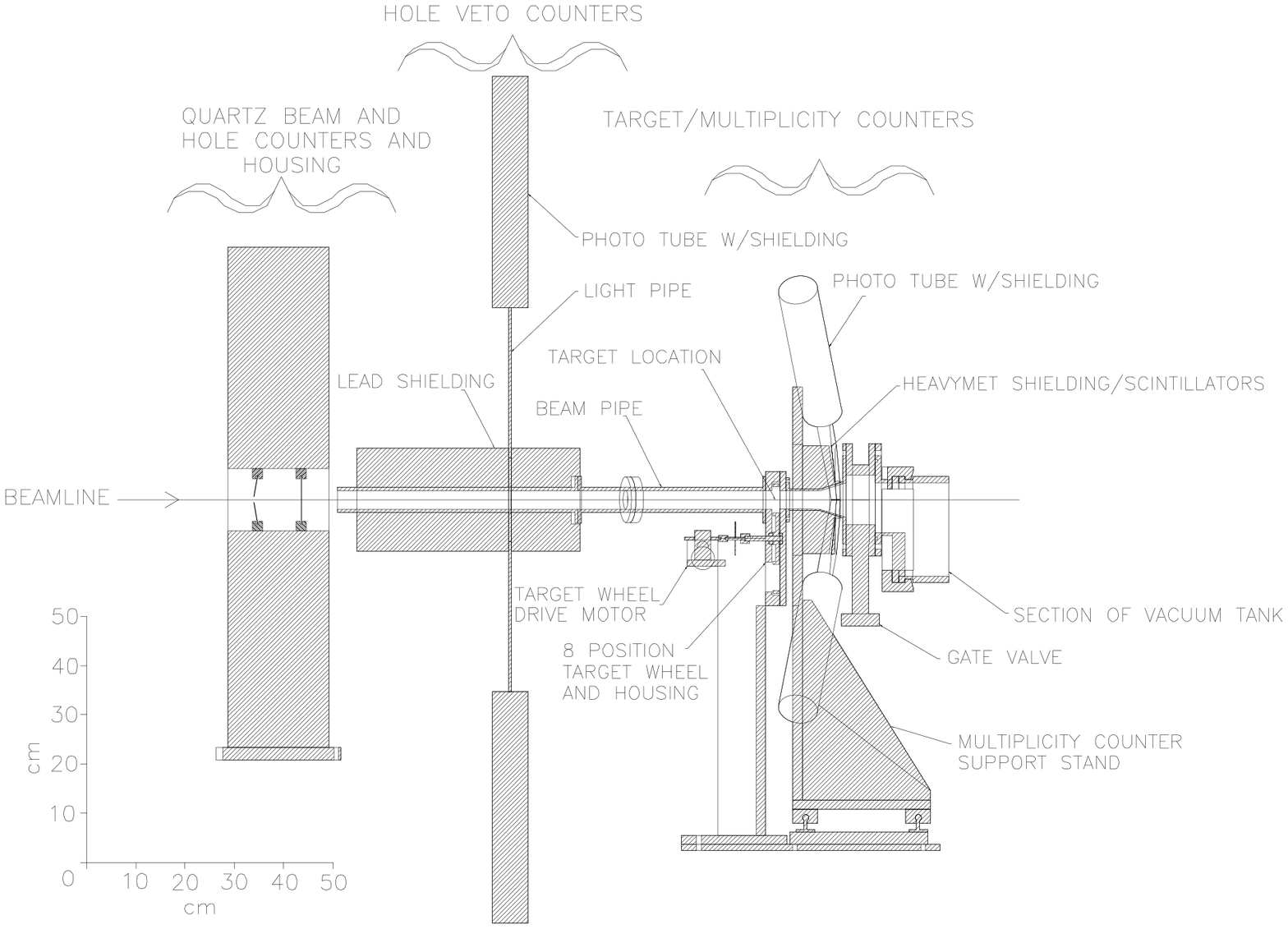}} 
\vskip 1.0 cm
\caption{The E864 front-end configuration of detectors consisting of the
Quartz-Cherenkov hole-veto and beam counters on the left, a scintillation
hole-veto counter in the middle followed by the target assembly and 
Multiplicity-trigger counter on the right.}
\label{horiz}
\end{center}
\end{figure}
\vskip 1.0 cm

\newpage
\begin{figure}[htb]
\begin{center}
\vskip -3.0 cm
\mbox{\epsfysize=14.cm\epsffile{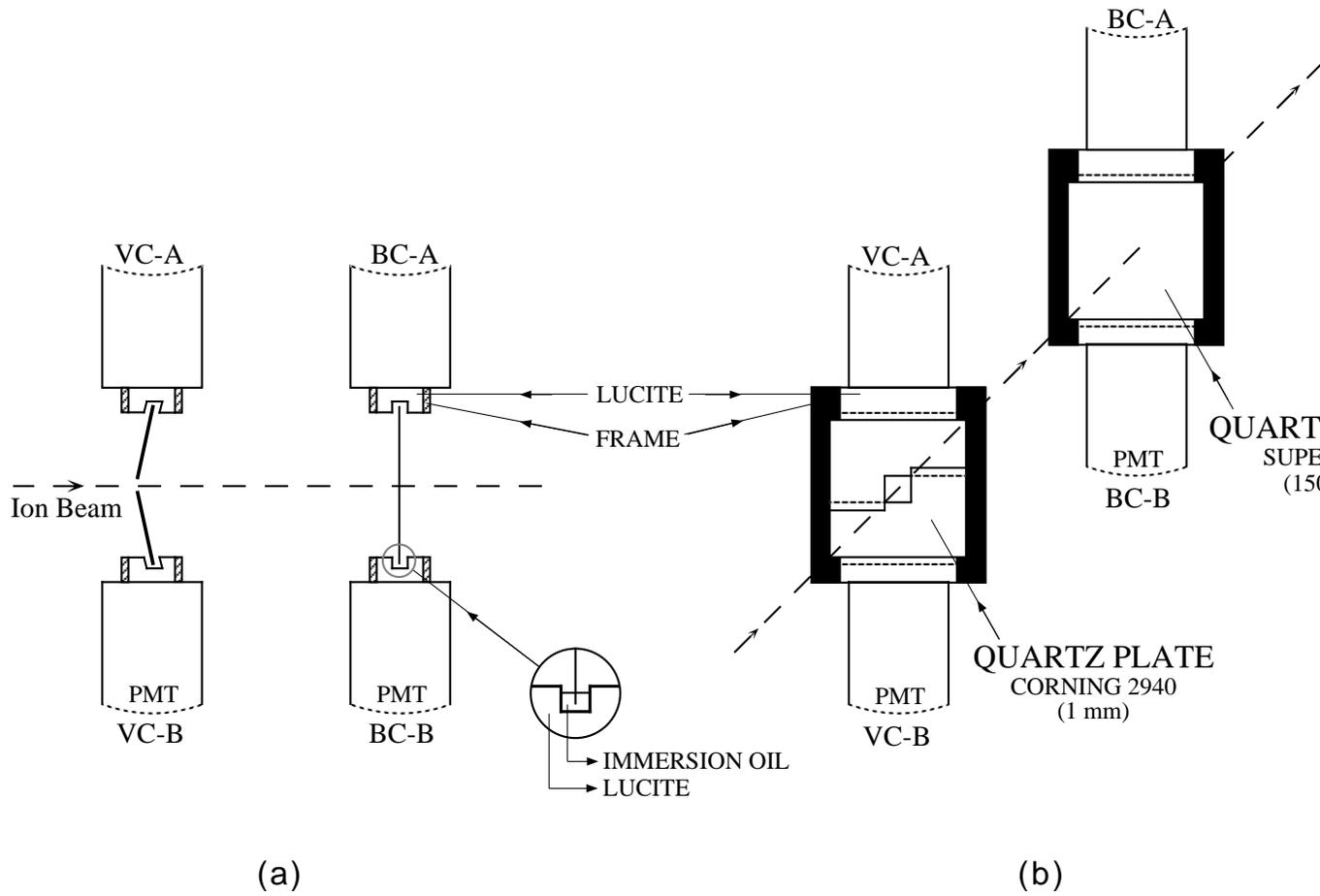}} 
\vskip 1.0 cm
\caption{A sketch illustrating the optical assembly of the quartz
hole-veto and beam counters; (a) shows the side view and (b) the perspective
view.}
\label{horiz}
\end{center}
\end{figure}
\vskip 4.0 cm

\newpage
\begin{figure}[htb]
\begin{center}
\vskip -2.5 cm
\mbox{\epsfxsize=14cm\epsffile{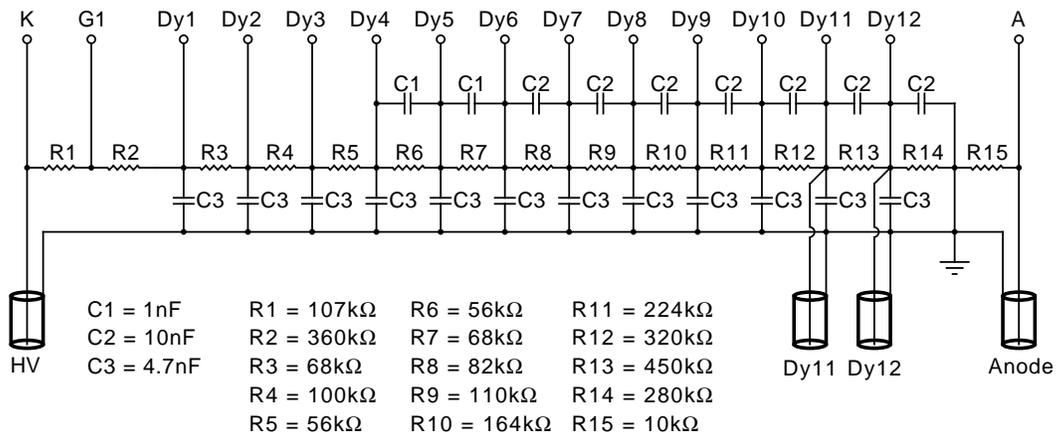}} 
\vskip 1.0 cm
\caption{The modified base for the Hamamatsu 1828-01 PMT with power
supplies attached to the last two dynodes.}
\label{horiz}
\end{center}
\end{figure}
\vskip 5.0 cm

\newpage
\begin{figure}[htb]
\vskip -2.0 cm
\mbox{\epsffile{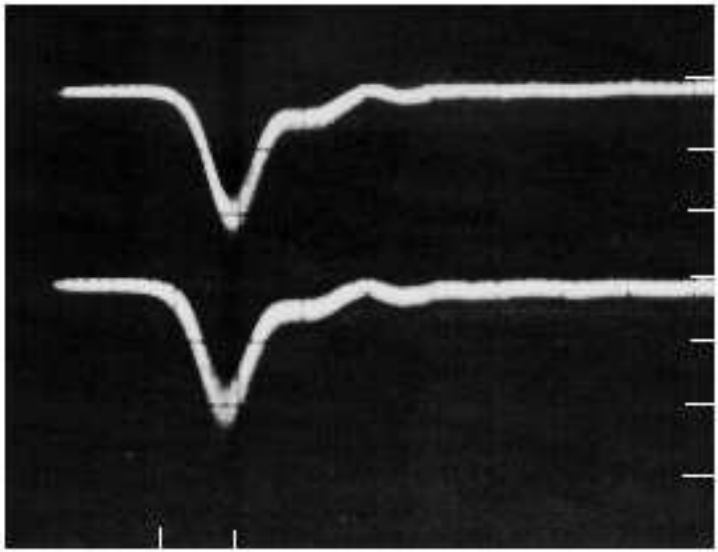}} 
\vskip 2.0 cm
\caption{Beam counter pulses during a spill of 1$\times$10$^{7}$ Au ions.
The vertical scale is 20~mV/div and the horizontal scale is 5~ns/div.}
\label{horiz}
\end{figure}
\vskip 1.0 cm

\newpage
\figure
\begin{center}
\PSbox{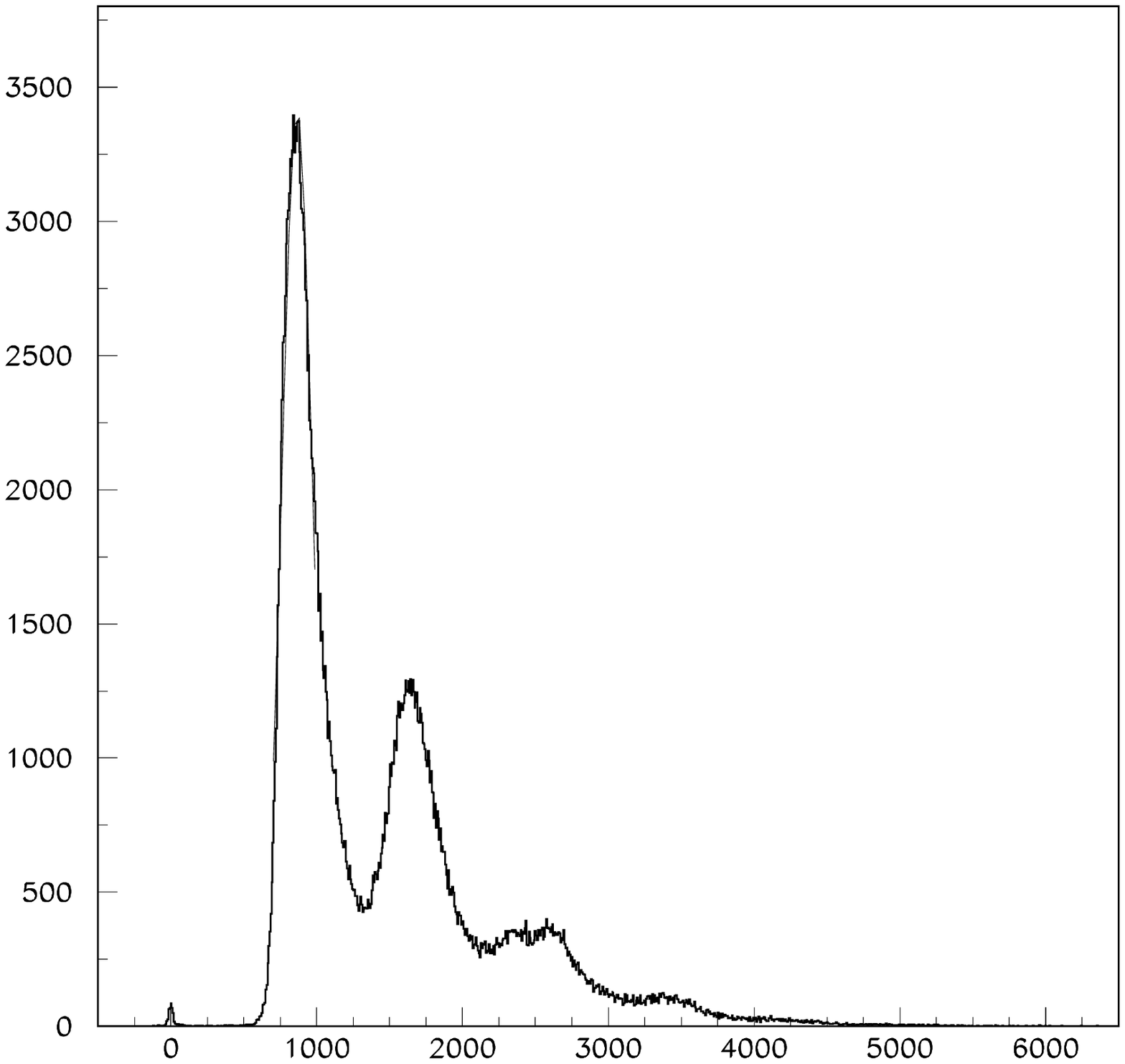 hoffset=-100 voffset=-288 hscale=90 vscale=90}
{4.0in}{3.75in}
\vskip 4.5cm
\centerline{\large\bf ADC CHANNEL NUMBER}
\vskip 2.0cm
\caption{The pulse height spectrum of the beam counter at an incident beam
rate of 1.2$\times$10$^{7}$ Au ions per second.}
\end{center}
\endfigure
\vskip 1.0cm

\newpage
\figure
\begin{center}
\PSbox{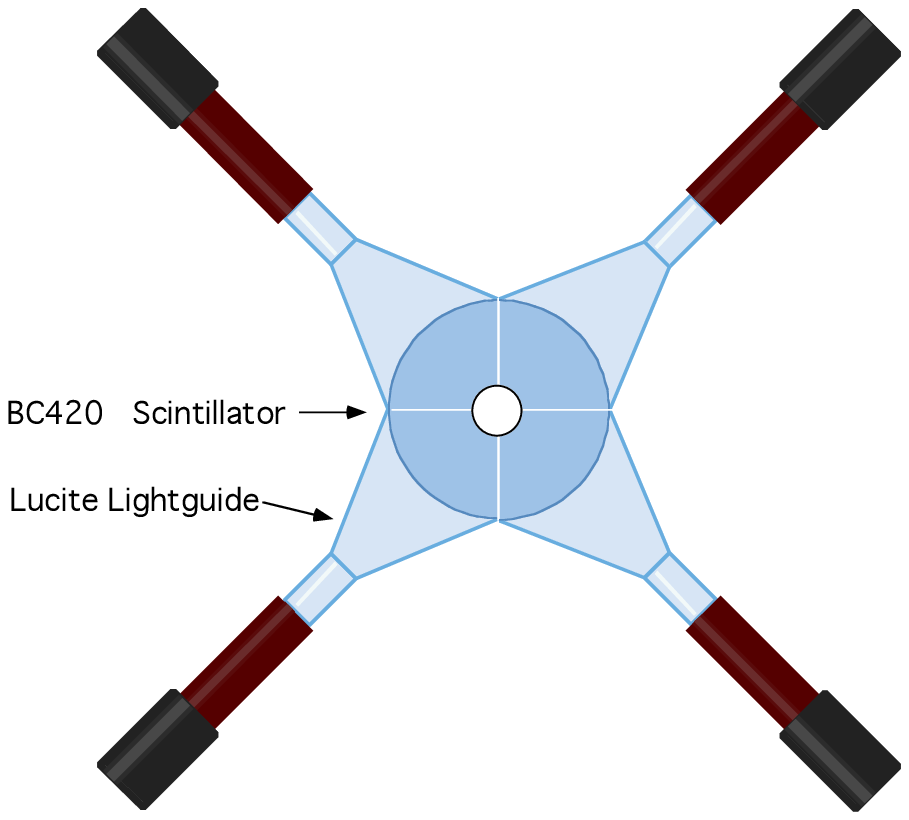 hoffset=-180 voffset=-270}{3.5in}{3.5in}
\vskip 1.0cm
\caption{Beam's eye view of the Multiplicity-trigger counters.}
\end{center}
\endfigure
\vskip 1.0cm

\newpage
\figure
\begin{center}
\PSbox{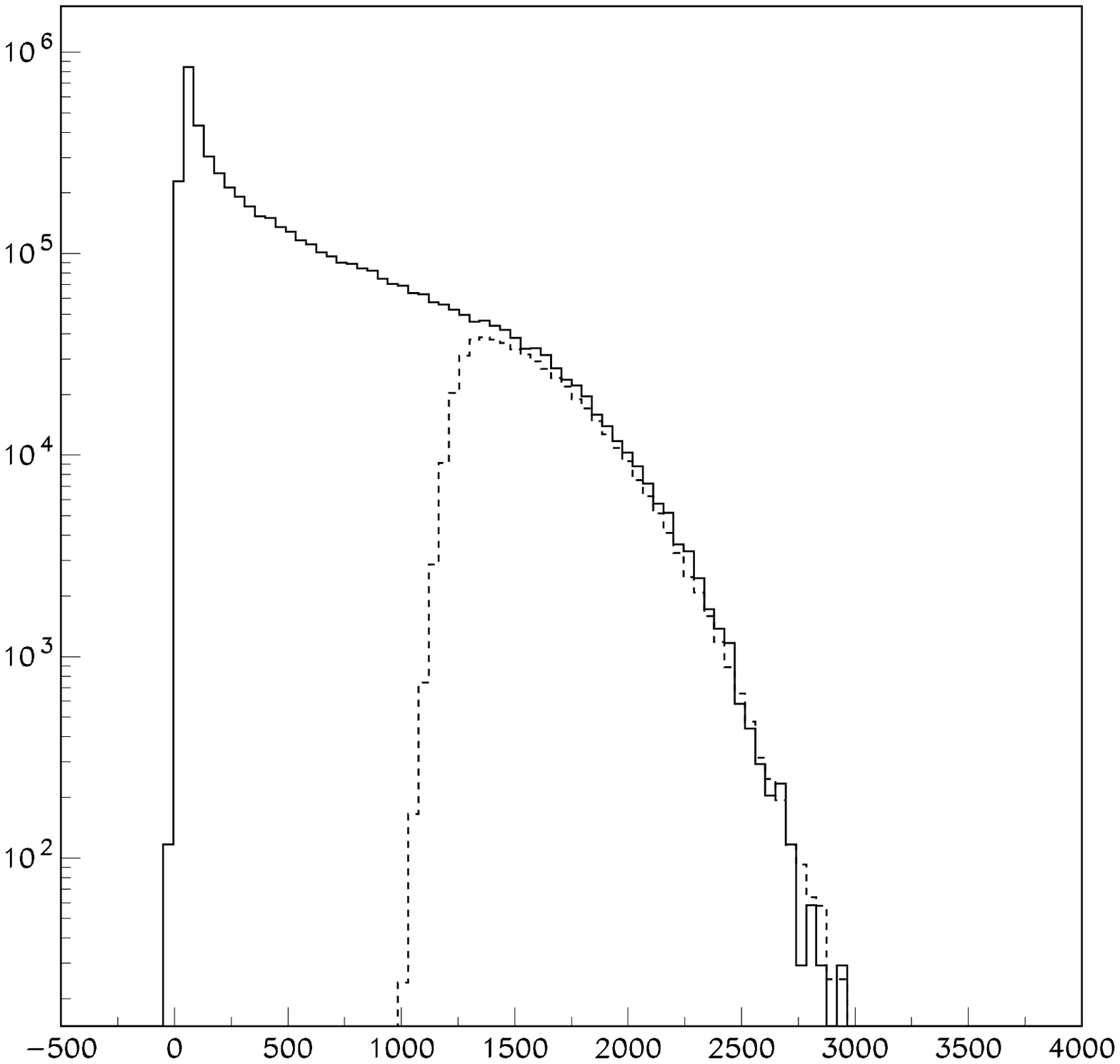 hoffset=-100 voffset=-288 hscale=90 vscale=90}
{4.0in}{3.75in}
\vskip 4.5cm
\centerline{\large\bf ~~~~~~ADC CHANNEL NUMBER}
\vskip 2.0cm
\caption{The averaged (over the four counters) Multiplicity detector 
pulse height distribution for
minimum-bias (solid line) and 10\% central events (dashed line).}
\end{center}
\endfigure
\vskip 1.0cm

\newpage
\figure
\begin{center}
\PSbox{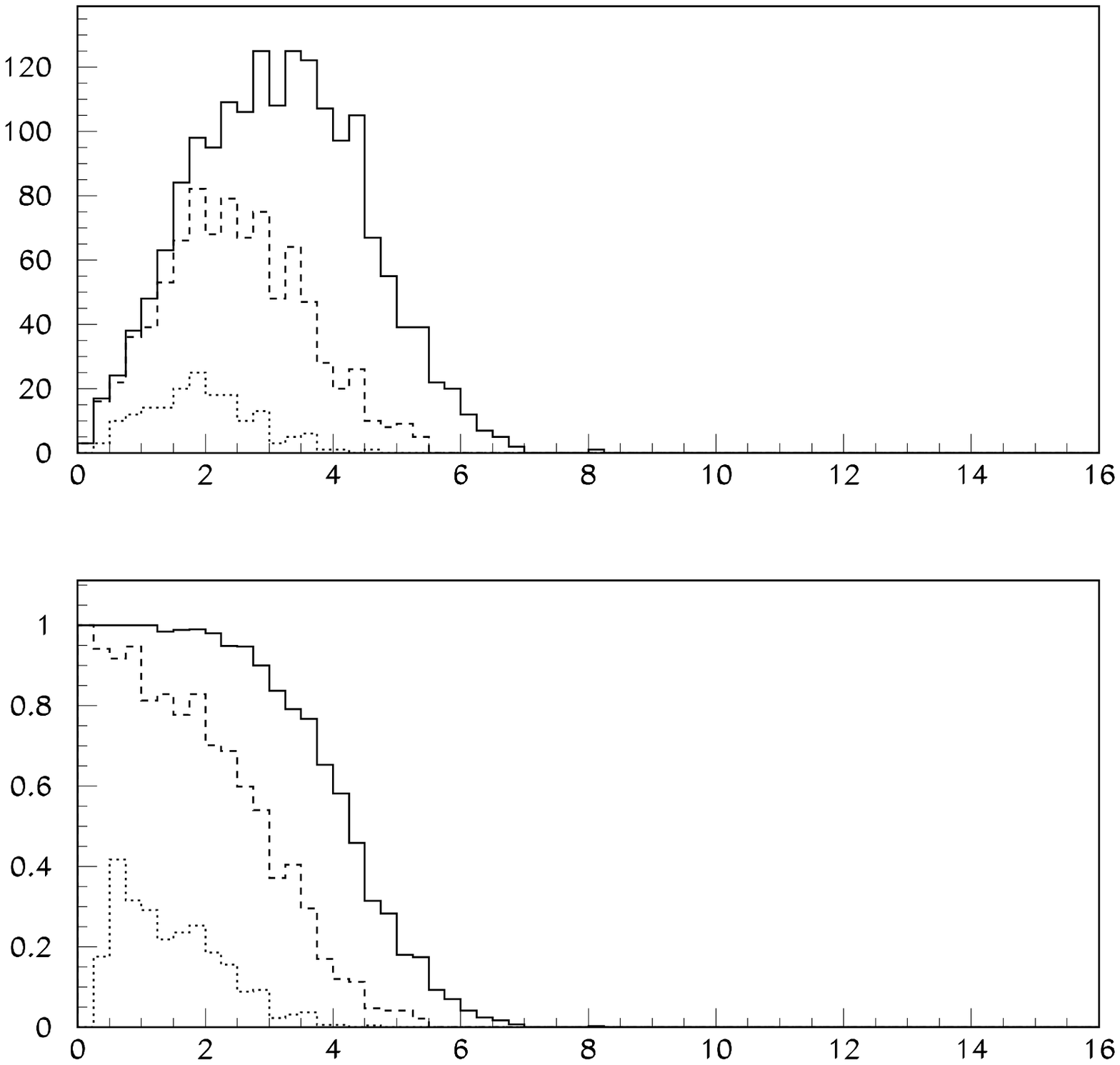 hoffset=-100 voffset=-288 hscale=90 vscale=90}
{4.0in}{3.75in}
\vskip 4.5cm
\centerline{\large\bf ~~~~~~~IMPACT PARAMETER OF COLLISION (fm).}
\vskip 1.0cm
\caption{Monte Carlo study of the centrality selection of the multiplicity
counter. The three different curves represent different cuts on the energy
deposited in the multiplicity counter. The solid line represents a cut on the
events with the 10\% most energy deposited, the dashed line represents the
5\% most energy deposited, and the dotted line represents the 1\% most
energy deposited. 
(a) Distribution of impact parameters (in fm) which pass a 10\%, 5\%
and 1\% cut on the energy deposited in the four scintillation counters in the
multiplicity counter. (b) Trigger probability for a particular impact parameter
event to pass a 10\%, 5\%, and 1\% energy deposit cut.}
\end{center}
\endfigure

\newpage
\figure
\begin{center}
\PSbox{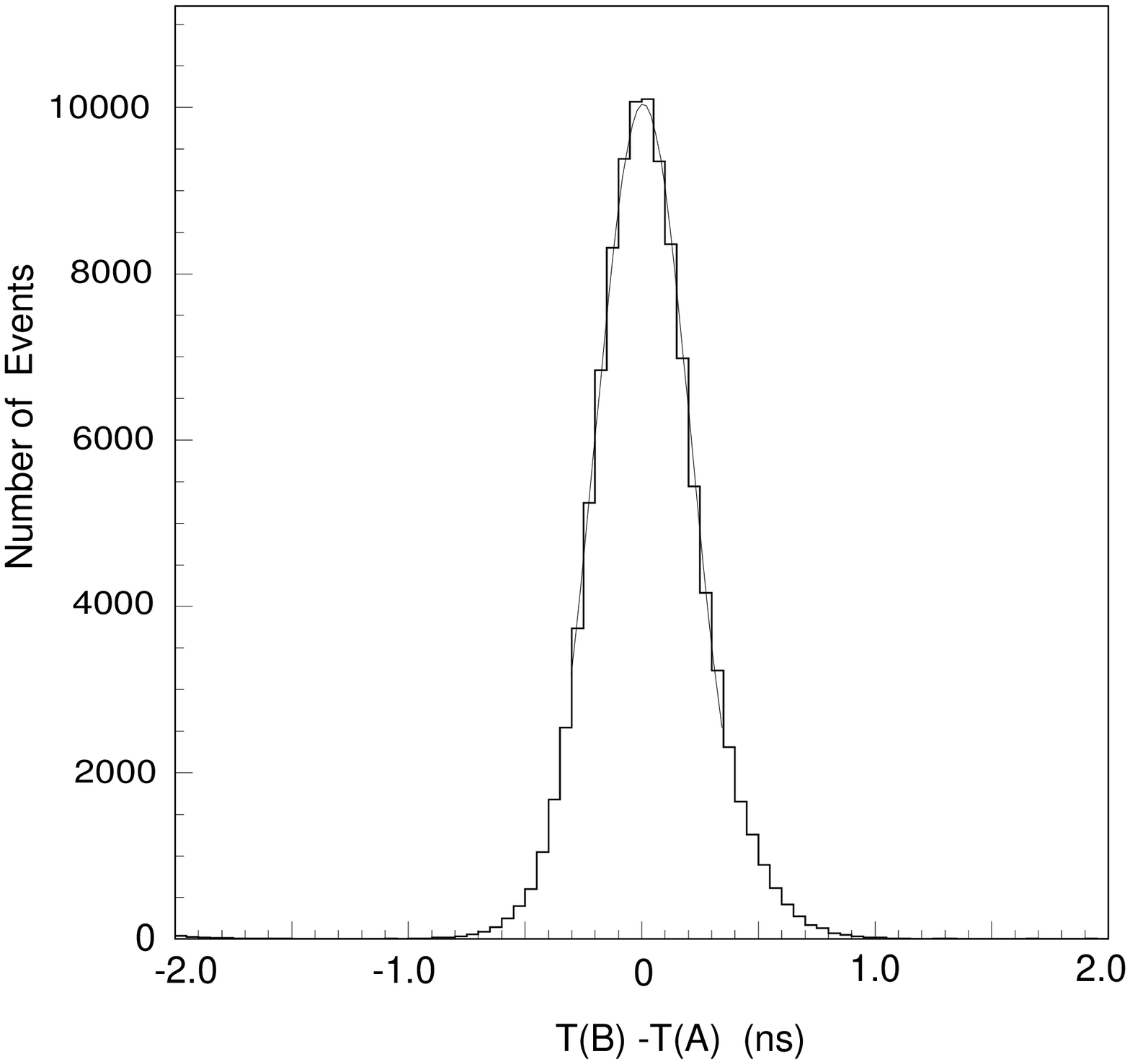 hoffset=-100 voffset=-288 hscale=90 vscale=90}
{4.0in}{3.75in}
\vskip 7.0cm
\caption{The distribution of the time difference between the bottom (B) and top
 (A) PMTs of the beam counter at an incident beam rate of 1$\times$10$^{7}$  
 Au ions. ~~~~$\sigma$(t$_{B}$-t$_{A}$)=200~ps and $\sigma$(t$_{0}$)=
$\sigma$(0.5(t$_{B}$+t$_{A}$)=100~ps.}
\end{center}
\endfigure

\newpage
\figure
\begin{center}
\PSbox{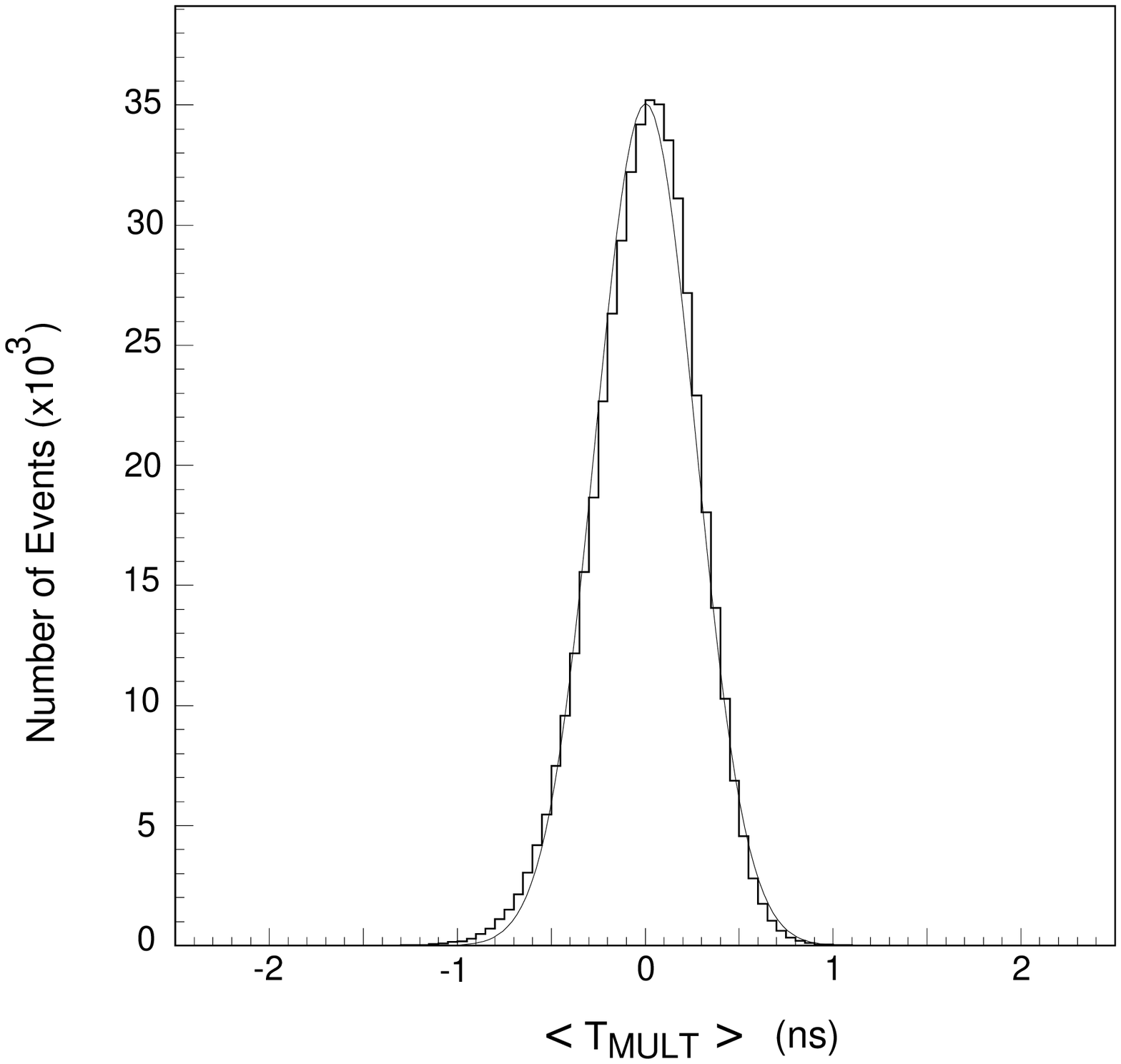 hoffset=-100 voffset=-288 hscale=90 vscale=90}
{4.0in}{3.75in}
\vskip 7.0cm
\caption{The averaged (over the four counters) intrinsic time resolution 
($\sigma$=250~ps) of the
Multiplicity-trigger counter system. }
\end{center}
\endfigure


\begin{thebibliography}{12}
\bibitem{4} Heavymet is the trade name for an alloy consisting of 95\% Tungsten,
            3.5\% Nickel and 1.5\% Copper and having a density 18.5~gm/cc.
\bibitem{7} Other tests of quartz Cherenkov counters have used much larger
            pulse heights (example 800mV, Private Communication, 
            Dana Beavis, BNL)
\end{thebibliography}
\end{document}